\documentclass[%
 reprint,
 amsmath,amssymb,
 aps,
]{revtex4-2}
\usepackage{amsmath}
\usepackage{graphicx}
\usepackage{dcolumn}
\usepackage{bm}
\usepackage{hyperref}
\usepackage{footnote}
\hypersetup{colorlinks=true, citecolor=blue, urlcolor=blue, linkcolor=blue}
\usepackage{subfigure}%
\usepackage{mathtools}  
\usepackage{subfigure}

\begin{document}


\title{Collective Nuclear Polaritons with Coherent and Tunable Excitation Dynamics}

\author{Liufeng Yang$^{1}$}%
\author{Jinling Wang$^{2}$}
\author{Huijun Li$^{1,2}$}
\email{hjli@zjnu.cn}
\author{Junhui Cao$^{3}$}%
\email{tsao.c@mipt.ru}
\author{Alexey Kavokin$^{3,4}$}%
\email{a.kavokin@westlake.edu.cn}
\author{Congjun Wu$^{4}$}%
\email{wucongjun@westlake.edu.cn}

\affiliation{$^{1}$Institute of Nonlinear Physics and Department of Physics, Zhejiang Normal University, Jinhua, 321004 Zhejiang, China\\
$^{2}$Zhejiang Institute of Photoelectronics, Jinhua, Zhejiang 321004, China\\
$^{3}$Abrikosov Center for Theoretical Physics, Moscow Center for Advanced Studies, Moscow 141701, Russia\\
$^{4}$School of Science, Westlake University, 18 Shilongshan Road, Hangzhou 310024, Zhejiang Province, China
}




\begin{abstract}
We propose collective nuclear polaritons formed by hybridizing a $^{229}\text{Th}$ nuclear ensemble with a vacuum-ultraviolet cavity mode generated via four-wave mixing, achieving a collective light-matter coupling that scales as $\sqrt{N}$. In the strong-coupling regime the system displays vacuum Rabi oscillations, indicating the hybridization between cavity photons and nuclear excitations. In the superradiant regime, the stored excitation is released in a cooperative burst with peak intensity scaling as $N^2$. The emission lifetime shrinks from thousands of seconds to the millisecond scale and remains tunable. Detuning sweeps across the polariton avoided crossing allow adiabatic conversion of the photonic excitation into a collective nuclear excitation, enabling reversible quantum storage. Our results demonstrate that cavity-mediated nuclear polaritons enable deterministic lifetime engineering and coherent quantum storage in nuclear systems.

\end{abstract}

\maketitle

\paragraph*{Introduction ---}

The isomeric transition of $^{229}\text{Th}$ is the lowest-energy nuclear transition known, with an excitation energy of about 8.4 eV that allows direct coherent driving by vacuum-ultraviolet (VUV) lasers \cite{PhysRevC.49.1845,MATINYAN1998199,2003Properties,tkalya2015radiative,PhysRevA.98.062520}. This unique property makes $^{229}\text{Th}$ an attractive candidate for nuclear clocks and fundamental physics tests \cite{PhysRevLett.108.120802,2019Energy,E.Peik_2003,xiao2026continuous,girvin2025prospectssolidstatenuclearclock}. Recent experiments have demonstrated direct laser excitation of the transition in solid-state environments and enabled high-precision measurements of its resonance frequency and lifetime \cite{PhysRevLett.133.013201,tiedau2024laser}. However, coherent control is hindered by the weak nuclear dipole moment and the long radiative lifetime $(\sim10^3~\mathrm{s})$ . Accessing the coherent regime requires enhancing the light-matter interaction strength beyond the intrinsic decay rate, a regime accessible via collective cavity quantum electrodynamics (QED) effects \cite{PhysRev.93.99,PhysRev.170.379}.

Controlling nuclear excitation dynamics while preserving coherence remains a central challenge. The collective cavity QED provides a route to this control, and has been widely exploited to enhance and control light-matter interactions in atomic, molecular, and solid-state systems \cite{RevModPhys.73.565,C.Weisbuch92,DavidLidzey98}. Under strong coupling, cavity photons hybridize with nuclear excitations to form polaritons, characterized by vacuum Rabi splitting and reversible energy exchange \cite{deng2010exciton}. Tuning the relative photonic and matter contributions, the lifetime and radiative properties of polaritons can be engineered.

Although strong coupling and coherent control have been observed in the X-ray regime, coherent nuclear polaritons in the laser-accessible VUV domain have not been demonstrated \cite{2016Collective,Liao_2012,Nickerson_2019}. The formation of nuclear polaritons offers a way to control excitation dynamics that goes beyond the limitations of free-space schemes. Exploiting the collective nature of the nuclear ensemble provides a $\sqrt{N}$ enhancement in coupling strength \cite{PhysRevLett.63.240}. Unlike purely dissipative Dicke superradiance, the strong-coupling regime supports hybridized light-matter quasiparticles  \cite{gross1982superradiance,PhysRevLett.30.309}. Through this hybridization, nuclear excitations acquire a photonic component that opens a controllable radiative decay channel. Tuning this hybridization controls the effective dissipation rate and enables reversible quantum storage. Analogously to polaritonic memories in atomic systems, adiabatic detuning sweeps map VUV photons onto stationary collective nuclear excitations, extending solid-state quantum memory protocols into the nuclear domain\cite{PhysRevLett.84.5094,Kaviani_2013,SciPostPhys.14.6.167,PhysRevLett.109.197403}.

In this work, we theoretically study a cavity-coupled $^{229}\text{Th}$ nuclear ensemble driven by a coherent VUV field generated via four-wave mixing (FWM). A plausible experimental route toward the required vacuum-ultraviolet confinement is a fluoride-based microcavity architecture built around the $^{229}\text{Th}$ with CaF$_2$ crystal itself. Fluoride materials are among the few optical media that remain transparent deep into the far- and vacuum-ultraviolet, making CaF$_2$ cavity spacers together with fluoride distributed Bragg reflectors a natural materials platform for operation near the 148 nm nuclear transition. We therefore adopt a theoretically reasonable microcavity target with micrometer-scale cavity length and transverse confinement, corresponding to an effective mode volume $V_{\rm eff}\approx10^{-15}\ \rm m^{3}$. This parameter range is sufficient to raise the single-nucleus coupling to the 100 Hz scale and, when combined with the collective $N$ enhancement, enables access to the collective strong-coupling regime considered in the following. We emphasize that realizing such a VUV microcavity remains a demanding photonic-engineering task, with mirror loss, thin-film absorption, and surface quality likely to be the main practical limitations. The collective $\sqrt{N}$ enhancement overcomes the weak nuclear dipole moment and brings the system into the strong-coupling regime. In this regime, cavity photons hybridize with collective nuclear excitations to form nuclear polaritons, giving rise to vacuum Rabi oscillations and clear spectral avoided crossings. The resulting dynamical phase diagram reveals a crossover between reversible Rabi oscillations and cooperative superradiant emission. The collective interaction also enables control of the effective radiative lifetime beyond the fixed spontaneous decay rate, allowing continuous manipulation of the isomeric state. In the strong-coupling regime, coherent quantum storage is further realized through an adiabatic detuning sweep.


\paragraph*{Model ---}\label{II}


\begin{figure}
    \centering
    \includegraphics[width=1.0\linewidth]{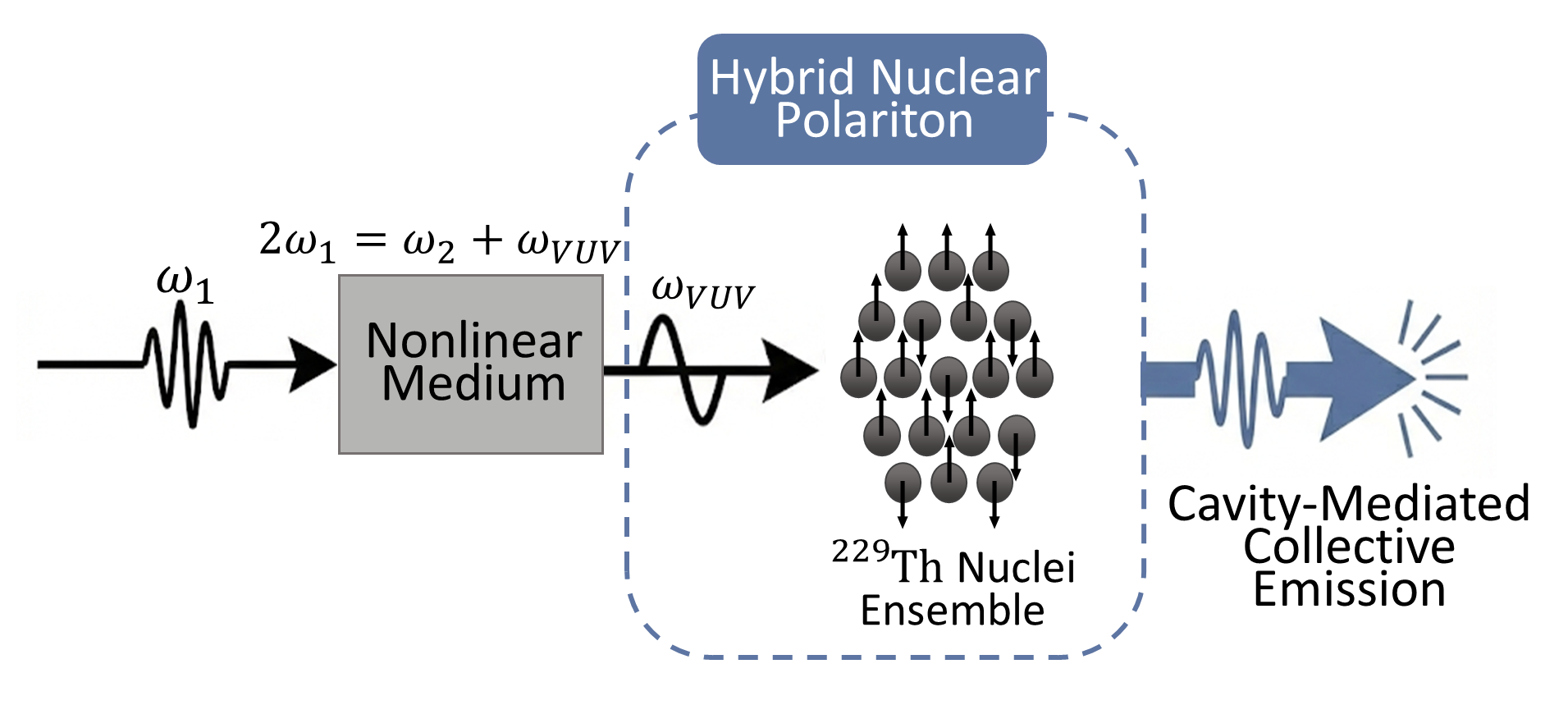}
    \caption{Schematic of collective nuclear polariton excitation. A pump field ($\omega_1$) enters a nonlinear medium and generates a VUV field via FWM ($2\omega_1 = \omega_2 + \omega_{\mathrm{VUV}}$, where $\omega_2$ is the idler field), which in turn drives a $^{229}\text{Th}$ ensemble to form hybrid nuclear polaritons. The system releases energy through a directional cavity-mediated collective emission.}\label{fig1}
\end{figure}

We consider an ensemble of $N$ two-level $^{229}\mathrm{Th}$ nuclei with transition frequency $\omega_{\mathrm{nuc}}$, collectively coupled to a VUV cavity mode of frequency $\omega_{\mathrm{VUV}}$, driven by a pump field ($\omega_1$) and a seed field ($\omega_2$). In the low-excitation regime $n_{\rm ex}=\Sigma_i\sigma^\dagger_i\sigma_i \ll N$, the collective nuclear spin operators can be bosonized using the Holstein-Primakoff transformation, $b^\dagger=\frac{1}{\sqrt N}\Sigma_i\sigma_i^\dagger$. The collective raising and lowering operators become $J_+=b^\dagger\sqrt{N-b^\dagger b}\approx \sqrt Nb^\dagger,J_-\approx \sqrt Nb$ and $J_z=b^\dagger b-N/2$.
The Hamiltonian of the system can be written as: 
\begin{equation}\label{H}
H = H_0 + H_\text{int}+ H_\text{FWM} + H_\mathrm{drive(t)}
\end{equation}
The first term governs the free evolution of the fields and the nuclear ensemble:
\begin{equation}
H_0 = \sum_{ j \in \{1,2,{\mathrm{VUV}}\}} \omega_{j} a_{j}^\dagger a_{j} + E_\text{nuc}J_z.
\label{eq:H0}
\end{equation}
The free Hamiltonian $H_0$, as defined in Eq.~(\ref{eq:H0}), governs the unperturbed evolution of the hybrid quantum system. The first term represents the energy of the multimode electromagnetic field, where $a^{\dagger}_{j}$ ($a_{j}$) denotes the bosonic creation (annihilation) operator for the $j$-th optical mode with eigenfrequency $\omega_\mathrm{j}$. These modes specifically include the fundamental pump fields ($j=1,2$) and the VUV cavity mode. The second term accounts for the energy of the nuclear ensemble, where $E_\text{nuc}$ corresponds to the nuclear transition energy of the $^{229}\text{Th}$ isomer.

The interaction between the VUV cavity mode and the nuclear ensemble is described by a Tavis-Cummings Hamiltonian: 
\begin{equation}
H_\text{int} = g ( a^{\dagger}_{\mathrm{VUV}} J_{-} + a_\mathrm{VUV} J_+ )
\end{equation}
where $g$ quantifies the single-photon coupling strength. The nonlinear interaction originates from the degenerate FWM process, which couples the pump, seed, and VUV modes: 
\begin{equation}
H_\text{FWM} = U ( a^{\dagger}_{\mathrm{VUV}} a^{\dagger}_{2} a^{2}_{1} + \rm h.c. )
\end{equation}
where $U$ denotes the FWM strength, satisfying $2\omega_1 = \omega_2 + \omega_{\mathrm{VUV}}$. In this scheme, the FWM process acts as a coherent source for the VUV cavity mode. The pump mode is coherently driven by a Gaussian pulse,
\begin{equation}
H_{\mathrm{drive(t)}}=
\Omega_p e^{-(t-t_0)^2/2\sigma^2}(a_1+a_1^\dagger).
\end{equation}
The proposed physical scheme is schematically illustrated in Fig.~\ref{fig1}.


The dissipative evolution of the coupled system is governed by the Lindblad master equation under the standard Born-Markov approximation \cite{Lindblad_1976,Breuer_2002}:
\begin{equation}
\partial_t \rho = -i [H, \rho] + \sum_{j\in\{1,2,\mathrm{VUV}\}} \kappa_{j}\mathcal{D}[a_{j}]\rho +  \gamma_-\mathcal{D}[J_-]\rho
\end{equation} 
Here, $\rho$ is the density operator of the system, and $H$ is the total Hamiltonian defined in Eq.~(\ref{H}). The non-unitary dynamics are described by the standard dissipator $\mathcal{D}[O]\rho=O\rho O^{\dagger}-\frac{1}{2}\{O^{\dagger}O,\rho\}$. The first summation accounts for photon losses in the optical modes ($j \in \{1,2,\mathrm{VUV}\}$) with decay rates $\kappa_j$, while the second term describes the intrinsic spontaneous emission of individual $^{229} \text{Th}$ nuclei with rate $\gamma_-$. 

We truncate the Hilbert space to the relevant photon sectors and single-excitation subspace (see Supplemental Material (SM)) to numerically solve the system dynamics. The light-matter interaction is mediated by the magnetic-dipole (M1) isomeric transition of $^{229}\text{Th}$. The single-photon coupling strength is given by $g = |\mu_\mathrm{ge}| B_{\mathrm{vac}}/\hbar$, where $B_{\mathrm{vac}} = [\mu_0 \hbar \omega /(2 V_{\mathrm{eff}})]^{1/2}$ denotes the vacuum magnetic-field fluctuations of the cavity mode \cite{scully1997quantum,weisskopf1930berechnung}. The transition magnetic moment $|\mu_\mathrm{ge}|$, intrinsic to the nucleus, determines the vacuum spontaneous decay rate $\Gamma_{\mathrm{vac}}$ (note that $\gamma_- = \Gamma_{\mathrm{vac}}$), according to
\begin{equation}\label{eq7}
\Gamma_{\mathrm{vac}} = \frac{\mu_0 \omega^3}{3\pi \hbar c^3} |\mu_\mathrm{ge}|^2.
\end{equation}
Using Eq.~(\ref{eq7}), the coupling strength becomes:
\begin{equation}
g = \sqrt{\frac{3\pi c^3 \Gamma_{\mathrm{vac}}}{2\omega^2 V_{\mathrm{eff}}}} .
\end{equation}
Utilizing the recently measured resonance frequency $\omega = 2020.409(7)\,\mathrm{THz}$ ($\lambda \simeq 148.38\,\mathrm{nm}$) and radiative lifetime $\tau_{\mathrm{vac}} = \Gamma_{\mathrm{vac}}^{-1} = 1740(50)\,\mathrm{s}$, and assuming a theoretically acceptable mode volume $V_{\mathrm{eff}} \approx 1 \times 10^{-15}\ \mathrm{m}^3$, we estimate a single-nucleus coupling strength of $g \approx 106.8\,\mathrm{Hz}$.

Equation (\ref{H}) describes the competition between coherent light-matter coupling and dissipation. In the weak-coupling regime, cavity loss and nuclear spontaneous decay dominate, leading to overdamped photon- and nuclear-like modes. When the collective coupling exceeds both the cavity decay rate $\kappa_{\mathrm{VUV}}$ and the nuclear decoherence rate $\gamma_{-}$, the system enters the strong-coupling regime, where photonic and nuclear excitations hybridize into nuclear polaritons, resulting in coherent vacuum Rabi oscillations. 

\begin{figure}
    \centering
    \includegraphics[width=1.0\linewidth]{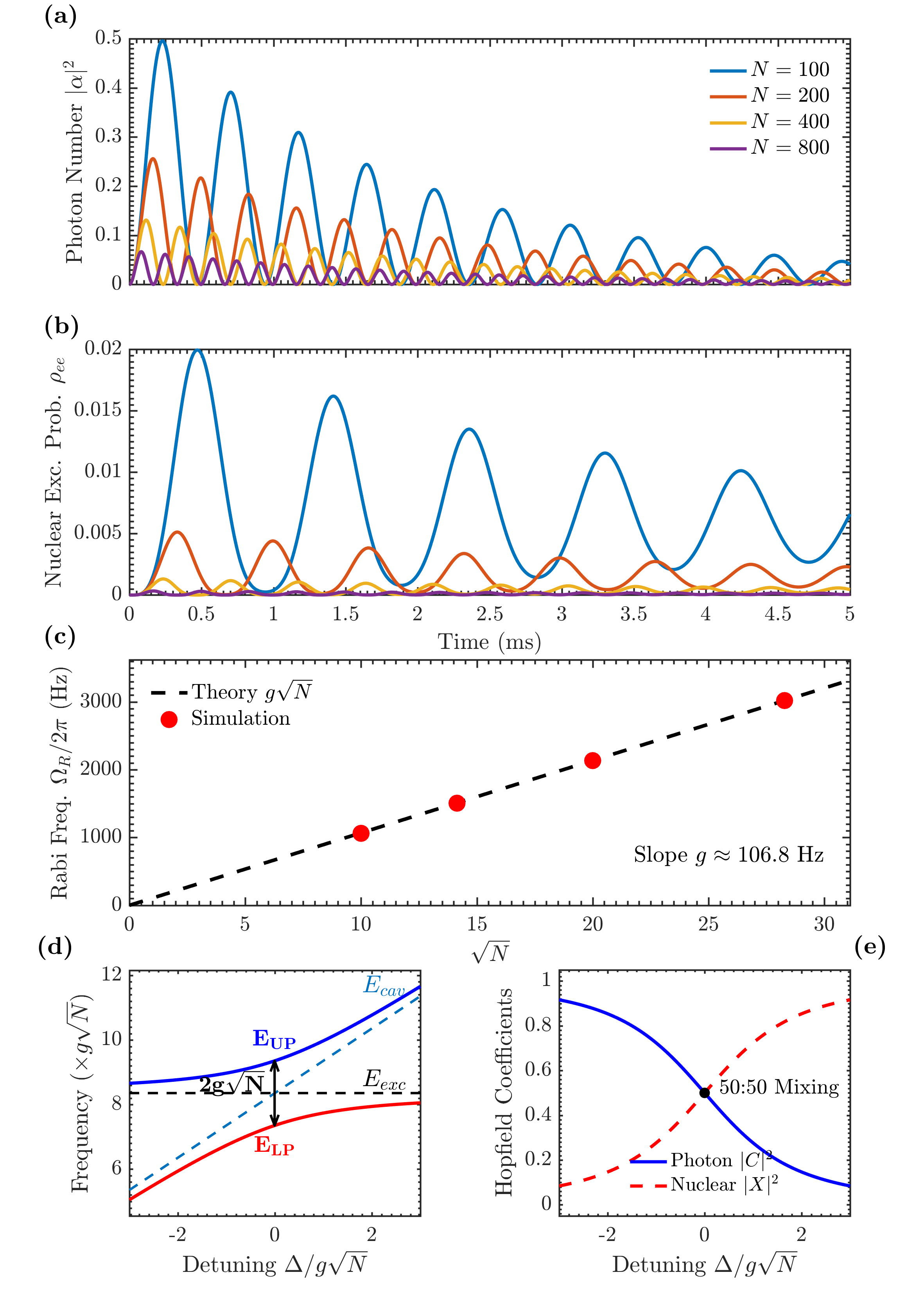}
    \caption{Dynamical and spectral signatures of collective nuclear polaritons. Time evolution of (a) the intracavity photon number $|\alpha|^2$ and (b) the nuclear excitation probability $\rho_{ee}$ for varying ensemble sizes $N$, illustrating vacuum Rabi oscillations. (c) Dependence of the Rabi frequency $\Omega_R/2\pi$ on $\sqrt{N}$. Red circles denote simulations; the dashed line represents the theoretical linear scaling $g\sqrt{N}$, giving a single-particle coupling strength of $g \approx 106.8$~Hz. The simulations assume a cavity decay $\kappa_{\mathrm{VUV}} = 1000$~Hz, fulfilling the strong-coupling criterion, with the corresponding working points marked in the phase diagram (Fig.~\ref{fig3}). (d) Eigenenergy spectrum versus normalized detuning $\Delta/g\sqrt N$. The avoided crossing between the upper (blue) and lower (red) branches shows a vacuum Rabi splitting of $2g\sqrt {N}$. Dashed lines denote the bare cavity and nuclear states.  (e) Photonic and nuclear fractions of the lower polariton. At resonance ($\Delta=0$), the photonic ($|C|^2$, solid blue) and nuclear ($|X|^2$, dashed red) components intersect at 0.5, indicating maximal light-matter hybridization.} \label{fig2}
\end{figure}

The collective dynamical evolution of the coupled system is governed by the mean-field expectation values of the relevant operators, 
\begin{equation} \alpha(t)=\langle\hat{a}\rangle,\quad  P(t)=\langle\hat{\sigma}_{-}^{(j)}\rangle,\quad Z(t)=\langle\hat{\sigma}_{z}^{(j)}\rangle,
\end{equation}
where $\alpha(t)$ denotes the coherent amplitude of the VUV cavity field, while $P(t)$ and $Z(t)$ represent the microscopic polarization and population inversion of a single nucleus, respectively.

We assume that all nuclei are identical and couple uniformly to the cavity mode. Using the mean-field approximation, $\langle\hat{a}\hat{\sigma}_-\rangle\approx\langle\hat{a}\rangle\langle\hat{\sigma}_-\rangle$, the Heisenberg equations of motion reduce to a closed set of nonlinear Maxwell-Bloch equations \cite{1072212},
\begin{equation}\label{eq10}
   \dot\alpha = \mathcal{E}(t) - \frac{\kappa_{\mathrm{VUV}}} {2}\alpha - iNgP,
\end{equation}
\begin{equation}\label{eq11}
\dot{P} = -\frac{\gamma_-} {2}P + ig\alpha Z,
\end{equation}
\begin{equation}\label{eq12}
\dot{Z} = -\gamma_-(Z+1) + 2ig(\alpha^* P - \alpha P^*) ,
\end{equation}
which self-consistently describe the interplay between coherent coupling, external driving, and dissipation.

Equation (\ref{eq10}) describes the intracavity VUV field driven by the effective FWM source $\mathcal{E}(t)$ and damped at the rate $\kappa_{\mathrm{VUV}}$, while $-iNgP$ captures the collective back-action of the nuclear ensemble. Eqs.~(\ref{eq11}) and ~(\ref{eq12}) govern the Bloch dynamics of the nuclei coupled to the cavity mode. The coupling terms $ig\alpha Z$ and $2ig(\alpha^{*}P-\alpha P^{*})$ account for coherent energy exchange between the cavity field and nuclear polarization, giving rise to vacuum Rabi oscillations when the collective coupling exceeds the dissipation rates.

In the strong-coupling regime [$g\sqrt{N} > (\kappa_\mathrm{VUV} + \gamma_-)/4$], Figs.~\ref{fig2}(a) and \ref{fig2}(b) depict the time evolution of the cavity field $|\alpha|^2$ and the nuclear excitation $\rho_\text{ee}$. The system exhibits clear vacuum Rabi oscillations associated with reversible energy exchange. The oscillation frequency increases with $N$, while the damping rate remains fixed by $\kappa_{\mathrm{VUV}}$, signaling the onset of collectively enhanced light-matter interactions. The extracted Rabi frequency $\Omega_\text{R}$ in Fig.~\ref{fig2}(c) follows the expected $\sqrt{N}$ scaling, yielding a single-nucleus coupling strength of $g \approx 106.8~\mathrm{Hz}$. This macroscopic scaling definitively confirms that the light-matter interaction is coherently enhanced by the nuclear ensemble.

The formation of nuclear polaritons is naturally described within the single-excitation subspace spanned by the bare VUV cavity photon state $|G,1_{\mathrm{VUV}}\rangle$ and the collective nuclear excited state $|E,0\rangle$ \cite{PhysRev.112.1555,C.Weisbuch92}. Within this subspace, the effective Hamiltonian reads
\begin{equation}
H_\text{eff}=   \begin{pmatrix}
\Delta&g\sqrt N \\g\sqrt N &0
\end{pmatrix},
\end{equation}
where $\Delta=\omega_{\mathrm{VUV}}-\omega_\text{nuc}$. The hallmark of strong coupling is the coherent and reversible exchange of energy between the cavity field and the nuclear ensemble at a rate exceeding the dissipative decay rates. To quantify the light-matter hybridization, we diagonalize $H_\text{eff}$ to obtain the lower ($|LP\rangle$) and upper ($|UP\rangle$) polariton states:
\begin{align}
|LP\rangle=C|G,1_{\mathrm{VUV}}\rangle-X|E,0\rangle,\\
|UP\rangle=X|G,1_{\mathrm{VUV}}\rangle-C|E,0\rangle ,
\end{align}
with $|C|^2+|X|^2=1$. The Hopfield coefficients $|C|^2$ and $|X|^2$ measure the photonic and nuclear fractions, respectively. Fig.~\ref{fig2}(d) and (e) illustrate the spectral signature of nuclear polariton formation. In Fig.~\ref{fig2}(d), the eigenenergy spectrum reveals a distinct avoided crossing at zero detuning ($\Delta=0$). The vacuum Rabi splitting of $2g\sqrt{N}$ between the upper (blue) and lower (red) polariton branches stands in sharp contrast to the intersecting bare modes (dashed lines). This splitting, which being larger than the relevant dissipative linewidth, indicates the onset of the collective strong coupling regime and the formation of hybridized nuclear polaritons. The hybrid nature of these eigen states is quantified by the Hopfield coefficients in Fig.~\ref{fig2}(e). At resonance, the equal admixture of photonic and nuclear components ($|C|^2 =|X|^2 = 0.5$) indicates coherent light-matter hybridization.

Beyond spectral hybridization, the formation of nuclear polaritons provides a controlled route to tuning the dissipative dynamics of the nuclear ensemble. In the strong-coupling regime, nuclear excitations acquire a finite photonic component, thereby opening an additional radiative decay channel mediated by cavity leakage. Importantly, a photonic admixture alone does not guarantee superradiance. Hybridization merely modifies the single-particle lifetime, whereas cooperative emission strictly requires macroscopic phase coherence \cite{PhysRevB.81.245419,PhysRevX.6.011025,Kirton2019}. To identify the regimes of these collective behaviors, we construct a phase diagram that establishes clear boundaries between the strong-coupling, superradiant, and weak-coupling regimes. It maps the tunable crossover from intrinsically inhibited decay to superradiant emission. 

\begin{figure}
    \centering
    \includegraphics[width=1\linewidth]{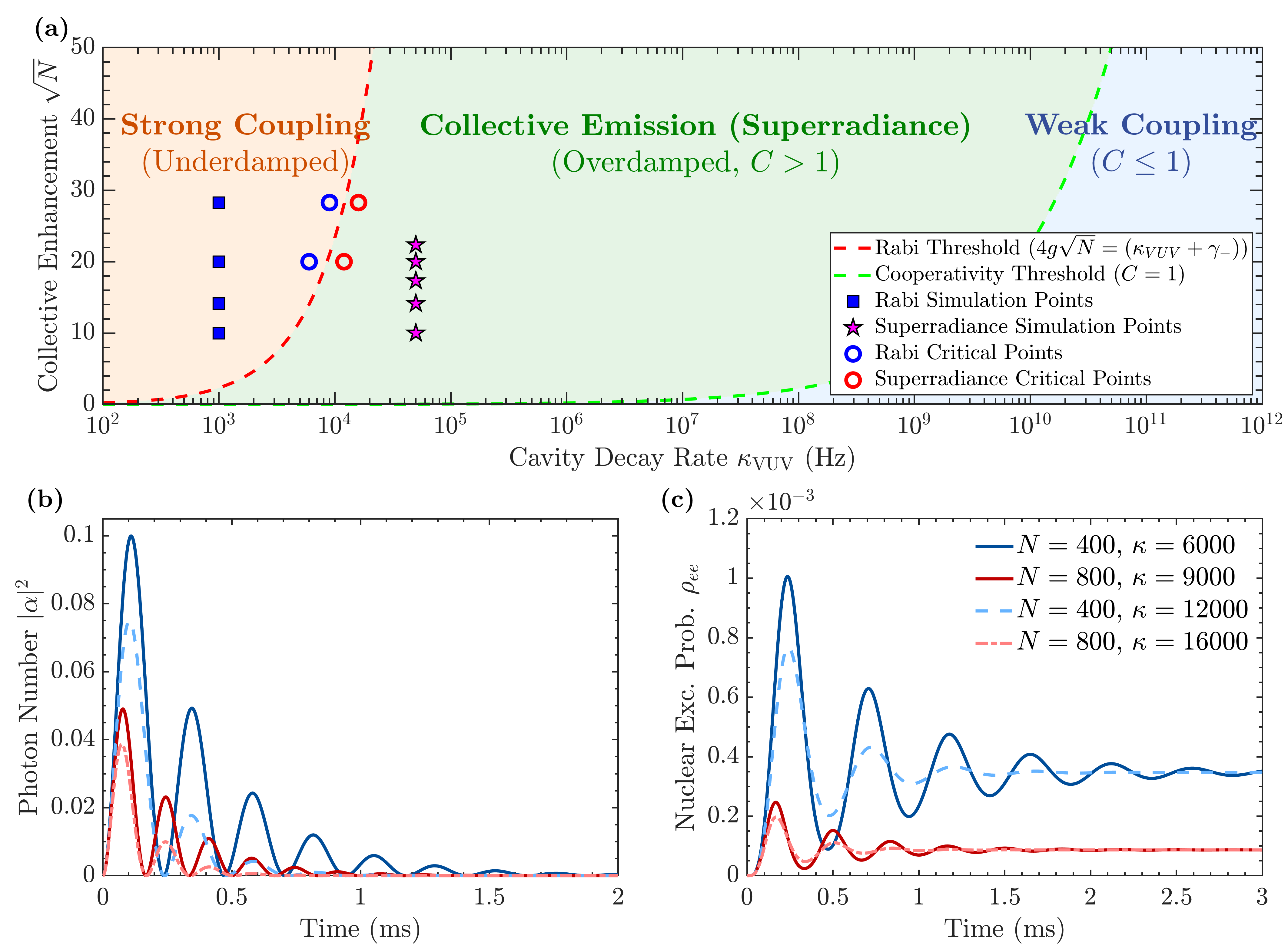}
    \caption{Collective phase diagram and critical dynamics of the cavity-nuclear ensemble. (a) Phase diagram as a function of the cavity decay rate $\kappa_\mathrm{VUV}$ (log scale) and the collective enhancement $\sqrt{N}$ (linear scale). The red dashed line denotes the strong-coupling threshold $4 g \sqrt{N} = (\kappa_{\mathrm{VUV}} + \gamma_-)$, separating underdamped vacuum Rabi dynamics from overdamped behavior, while the green dashed line marks the cooperativity boundary $C=1$. Blue, green, and orange regions denote weak coupling, collective emission (superradiance), and strong coupling, respectively. Magenta pentagons and blue squares indicate the working points used in the superradiance and Rabi-oscillation simulations, while open circles mark the critical transition points. (b) Intracavity photon number $|\alpha|^2$ and (c) the nuclear excitation probability $\rho_{ee}$ dynamics  for fixed ensemble sizes $N=400$ (blue curves) and $N=800$ (red curves). Solid (dashed) curves correspond to evolution toward (away from) the superradiant regime across the strong-coupling boundary ($(\kappa_{\mathrm{VUV}}+\gamma_-) \leq 4g\sqrt{N}$).} \label{fig3}
\end{figure}

Fig.~\ref{fig3}(a) presents a comprehensive phase diagram of the cavity-nuclear ensemble system, elucidating the dynamical competition between coherent collective coupling and multi-channel dissipation. The system behavior depends on two parameters: the strong-coupling threshold ($4 g \sqrt{N} > (\kappa_{\mathrm{VUV}} + \gamma_-)$) and the cooperativity boundary ($C=Ng^2/(\kappa_{\mathrm{VUV}}\gamma_-)=1$) \cite{HJKimble_1998,PhysRevA.81.033847}. These parameters define the boundary between the single-atom weak-coupling regime and collective superradiance.

In the strong-coupling regime, the collective vacuum Rabi frequency exceeds the total decoherence arising from both cavity leakage and intrinsic nuclear decay, leading to underdamped coherent energy exchange. The blue squares indicate the parameters used to demonstrate Rabi oscillations in Fig.~\ref{fig2} (b), confirming their location deep within the robust strong-coupling domain.

In the regime defined by $C > 1$ with the overdamped limit ($(\kappa_{\mathrm{VUV}}+\gamma_-) > 4 g \sqrt{N}$, green shaded area), the cavity mode acts as an effective dissipative channel that mediates collective nuclear decay. The magenta pentagons correspond to the superradiance simulations shown in SM Fig.~(S1), validating the emergence of collective emission in the bad-cavity limit.

The blue and red open circles identify the critical points at which the system transitions between coherent nutation and dissipative burst dynamics. Delineating these phase boundaries validates the collective $\sqrt{N}$ scaling. This shows that tuning the ensemble size $N$ and cavity decay rate $\kappa_{\mathrm{VUV}}$ provides direct control over the nuclear polariton dynamics. 

To further characterize this collective emission, we focus on the emission dynamics in the bad-cavity limit ($\kappa_{\mathrm{VUV}} \gg g\sqrt{N}$). Here, the cavity field can be adiabatically eliminated, yielding a enhanced collective decay rate $\Gamma_{\mathrm{eff}}=\gamma_- + \frac{4Ng^2}{\kappa_{\mathrm{VUV}}}$. The resulting emission exhibits macroscopic phase coherence ($g^{(1)}(t) \rightarrow 1$) and obeys the Dicke superradiance scaling, with a peak intensity $I_{\mathrm{max}} \propto N^2$ (see SM Fig.~S1). In this regime, the radiative lifetime can be tuned continuously: the effective pulse duration scales linearly with the cavity decay rate ($\tau_{\mathrm{eff}}\sim \kappa_{\mathrm{VUV}}/(Ng^2)$), forming a tunable interface between the long-lived nuclear isomer and fast optical modes (see SM Fig.~S2).

Adiabatic elimination yields the correct $N^2$ scaling of superradiant emission in the bad-cavity limit, but fails to capture the buildup of coherent energy exchange due to the neglect of cavity-field dynamics. In the crossover regime where collective coupling approaches the cavity decay rate, the adiabatic approximation breaks down. Retaining the cavity mode and solving the full Maxwell-Bloch equations is therefore required to model the transition to coherent strong coupling. 

Figs. \ref{fig3}(b) and (c) show this dynamical crossover for fixed ensemble sizes ($N=400, 800$). Parameters slightly below the collective coupling threshold ($(\kappa_{\mathrm{VUV}}+\gamma_-) \leq 4g\sqrt{N}$) exhibit damped vacuum Rabi oscillations (solid lines). Above this threshold (dashed lines), cavity dissipation suppresses coherent exchange, demonstrating a direct transition from coherent oscillations to overdamped superradiance as cavity finesse changes.

\begin{figure}
    \centering
    \includegraphics[width=1.0\linewidth]{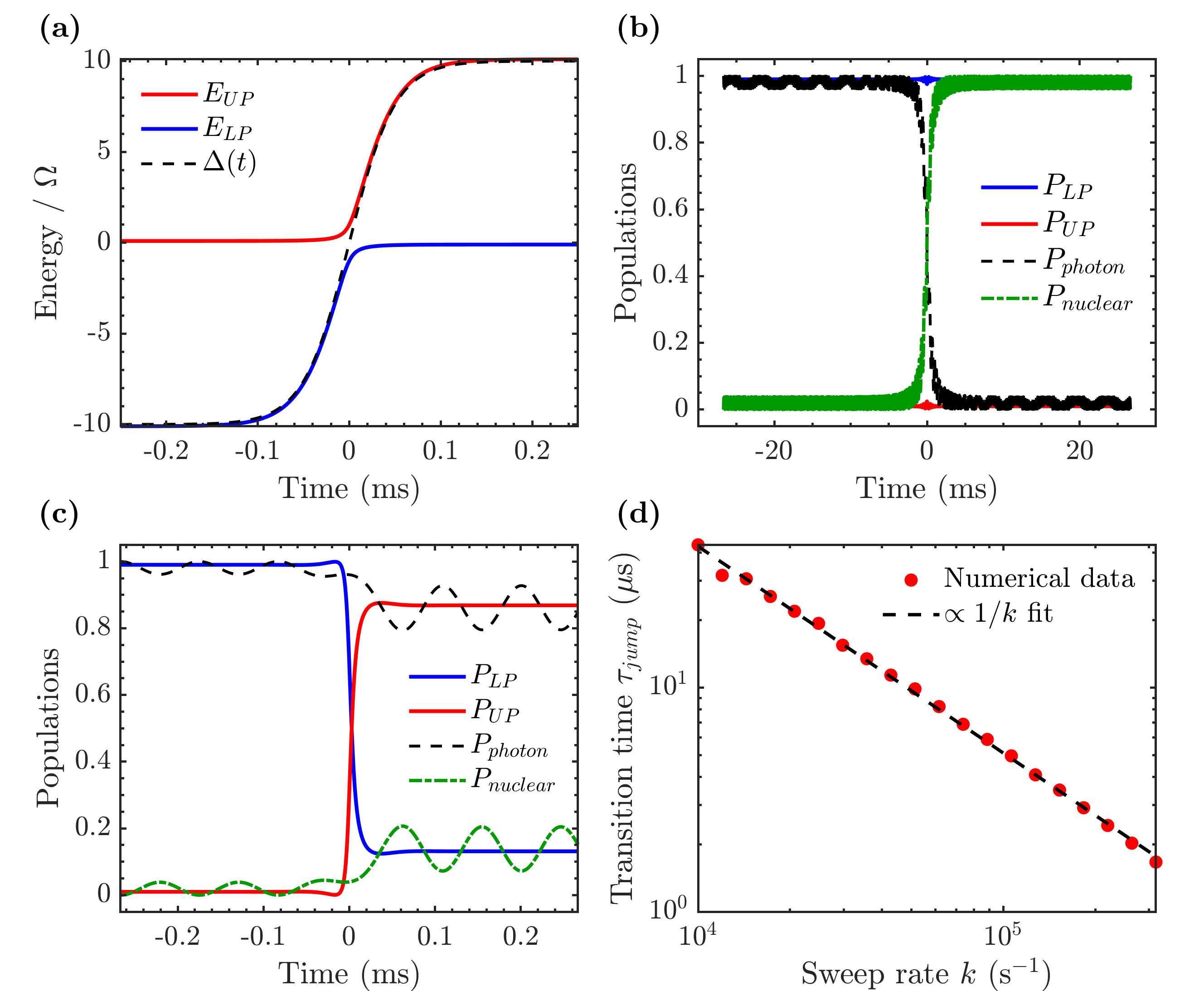}
    \caption{Coherent sweep dynamics of collective nuclear polaritons for an ensemble of $N=100$. (a) Instantaneous eigenenergies of the upper (UP) and lower (LP) polariton branches under the detuning sweep $\Delta(t)=\Delta_{0}\tanh(kt)$. The dashed curve denotes the bare photonic energy $\Delta(t)$. An avoided crossing with minimum gap $2\Omega$ appears at resonance. (b) Adiabatic regime ($k=300~\mathrm{s}^{-1}$). The system follows the LP branch, achieving near-complete photonic-to-nuclear conversion. (c) Nonadiabatic regime ($k=3.0\times10^{4}~\mathrm{s}^{-1}$). The fast sweep induces a diabatic transition to the UP branch, and the bare photonic population remains dominant after resonance. (d) Transition duration $\tau_{\mathrm{jump}}$ as a function of sweep rate $k$. The dashed line indicates the $k^{-1}$ scaling.} \label{fig4}
\end{figure}

Beyond lifetime engineering, the strong-coupling regime enables reversible quantum state transfer between photonic and nuclear degrees of freedom. We illustrate this capability through a quantum storage protocol. We study the coherent dynamics of collective nuclear polaritons under the detuning sweep $\Delta(t)=\Delta_{0}\tanh(kt)$ with $\Delta_{0}\gg\Omega$. In the limit $t\rightarrow -\infty$, the system is initialized in the bare photonic state $|C\rangle$. As shown in Fig.~\ref{fig4}(a), this state coincides with the lower polariton (LP) branch. The dynamics is governed by the avoided-crossing gap and the Landau-Zener parameter $\Gamma=\frac{\pi\Omega^{2}}{k\Delta_{0}}$ \cite{Zener1932}.

In the adiabatic regime ($\Gamma\gg1$), the system follows the LP branch [Fig.~\ref{fig4}(b)]. In the dressed-state picture, the LP character evolves from photonic to nuclear, leading to near-complete excitation transfer ($P_{\mathrm{nuclear}}\rightarrow1$). For a rapid sweep ($\Gamma\ll1$), Fig.~\ref{fig4}(c) shows that the system undergoes adiabatic transition from the LP to the upper polariton (UP) branch at the avoided crossing. The excitation therefore preserves its photonic nature and displays persistent quantum beating beyond resonance, with an asymptotic frequency approaching $\Delta_{0}$.

The adiabatic jump is characterized by the transition time window $\tau_{\mathrm{jump}} = t_{90}-t_{10}$. In Fig.~\ref{fig4}(d), the numerical results follow the scaling $\tau_{\mathrm{jump}} \propto k^{-1}$, over two orders of magnitude in the sweep rate, confirming that the non-adiabatic conversion is temporally localized in the vicinity of the resonance.

\paragraph*{Conclusion ---}

Collective nuclear polaritons provide a versatile platform for the coherent control of the $^{229}\text{Th}$ isomeric transition, bridging the disparate timescales of nuclear physics and quantum optics. The interplay between coherent FWM and cavity-mediated dissipation enables a dynamical reconfiguration of the nuclear ensemble, allowing a controlled crossover from reversible vacuum Rabi oscillations to the collective emission. This capability to continuously tune the radiative lifetime elevates the nuclear ensemble from a passive frequency reference to an active quantum medium. Dynamic control of the polariton branches dictates a reversible conversion between VUV photons and nuclear excitations. The present framework establishes a theoretical foundation for next-generation nuclear lattice clocks with reduced interaction-induced shifts, as well as for the realization of phase-coherent vacuum-ultraviolet light sources, and solid-state quantum memories.

AVK acknowledges support from Saint Petersburg State University (Research Grant No. 125022803069-4) and from the Innovation Program for Quantum Science and Technology (No. 2021ZD0302704). CW  acknowledges National Natural Science Foundation of China (Grants No. 12574274, No. 12234016, and No. 12550402) and the New Cornerstone Science Foundation. HJL acknowledges Natural Science Foundation of Zhejiang Province of China (Grants No. LZ26A040002 and No. LJHSQY26A040002) and National Natural Science Foundation of China (Grant No. 12375006).

LY and JW contributed equally to this work.


\bibliography{apssamp}

\end{document}


\title{Supplementary Information}

\author{Liufeng Yang$^{1}$}%
\author{Jinling Wang$^{2}$}
\author{Huijun Li$^{1,2}$}
\email{hjli@zjnu.cn}
\author{Junhui Cao$^{3}$}%
\email{tsao.c@mipt.ru}
\author{Alexey Kavokin$^{3,4}$}%
\email{a.kavokin@westlake.edu.cn}
\author{Congjun Wu$^{4}$}%
\email{wucongjun@westlake.edu.cn}

\affiliation{$^{1}$Institute of Nonlinear Physics and Department of Physics, Zhejiang Normal University, Jinhua, 321004 Zhejiang, China\\
$^{2}$Zhejiang Institute of Photoelectronics, Jinhua, Zhejiang 321004, China\\
$^{3}$Abrikosov Center for Theoretical Physics, Moscow Center for Advanced Studies, Moscow 141701, Russia\\
$^{4}$School of Science, Westlake University, 18 Shilongshan Road, Hangzhou 310024, Zhejiang Province, China
}




\begin{abstract}
           
\end{abstract}

\maketitle

\section{Hamiltonian matrix and basis vectors}\label{I}

Consider a three-mode system (optical modes $a_1, a_2, a_{\mathrm{VUV}}$) coupled to $N$ two-level atoms. The system Hamiltonian reads:
\begin{equation}
\begin{aligned}
H &= H_0 + H_{\rm in} + H_{\rm FWM} + H_{\rm pump} \\
&= \omega_1 a_1^\dagger a_1 + \omega_2 a_2^\dagger a_2 + \omega_{\mathrm{VUV}} a_{\mathrm{VUV}}^\dagger a_{\mathrm{VUV}} + E_{\rm ex} J_z \\
&\quad + g (a_{\mathrm{VUV}}^\dagger + a_{\mathrm{VUV}}) (J_+ + J_-) \\
&\quad + U \left( a_{\mathrm{VUV}}^\dagger a_2^\dagger a_1 a_1 + \text{h.c.} \right) \\
&\quad + \Omega_p e^{-(t-t_0)^2 / 2\sigma^2} (a_1 + a_1^\dagger),
\end{aligned}
\end{equation}
where $a_1$, $a_2$, and $a_{\mathrm{VUV}}$ are the annihilation operators for the optical field modes; $J_+$, $J_-$, and $J_z$ denote the collective atomic spin operators; $g$ is the atom-cavity coupling strength; $U$ characterizes the four-wave mixing (FWM) coupling strength; $\omega_p$ denotes the pump field amplitude; and $E_{\mathrm{ex}}$ represents the atomic transition energy.

We model the coupled dynamics within a truncated 11-dimensional Hilbert space spanned by the basis $\mathcal{B} = \{ |n_1,n_2,n_{\mathrm{VUV}},n_{\mathrm{nuc}}\rangle \}$. Here, $n_{1,2,\mathrm{VUV}}$ denote the photon numbers in the pump, seed, and VUV modes, respectively, while $n_{\mathrm{nuc}} \in \{0, 1\}$ represents the collective nuclear excitation. The explicitly ordered basis is:
\begin{equation}
\begin{aligned}
\mathcal{B} = \{ &|2,0,0,0\rangle, |0,1,1,0\rangle, |0,1,0,1\rangle, |1,0,0,0\rangle, \\
                 &|1,1,1,0\rangle, |0,0,1,0\rangle, |0,0,0,1\rangle, |1,0,1,0\rangle, \\
                 &|0,1,0,0\rangle, |1,0,0,1\rangle, |1,1,0,1\rangle \}.
\end{aligned}
\end{equation}
In the rotating frame, system evolution is driven by two primary off-diagonal mechanisms. The four-wave mixing (FWM) interaction $U$ governs multiphoton transitions under energy conservation (e.g., $|2,0,0,0\rangle \leftrightarrow |0,1,1,0\rangle$). Concurrently, the collective vacuum Rabi coupling $g$ hybridizes the bare VUV photon $|G, 1_{\mathrm{VUV}}\rangle \equiv |0,0,1,0\rangle$ with the single nuclear excitation $|E, 0\rangle \equiv |0,0,0,1\rangle$, driving the formation of collective nuclear polaritons.

The full matrix form of the Hamiltonian is given as follows:

\begin{widetext}
    \begin{equation}
[H = \left( {\begin{array}{*{20}{c}}
{2{\omega _1}}&{\sqrt 2 U}&0&{\sqrt 2 {\omega _p}}&0&0&0&0&0&0&0\\
{\sqrt 2 U}&{{\omega _2} + {\omega _{VUV}}}&g&0&{{\Omega _p}}&0&0&0&0&0&0\\
0&g&{{\omega _2} + {E_{ex}}}&0&0&0&0&0&0&0&{{\Omega _p}}\\
{\sqrt 2 {\Omega _p}}&0&0&{{\omega _1}}&0&0&0&0&0&0&0\\
0&{{\Omega _p}}&0&0&{{\omega _1} + {\omega _2} + {\omega _{VUV}}}&0&0&0&0&0&g\\
0&0&0&0&0&{{\omega _{VUV}}}&g&{{\Omega _p}}&0&0&0\\
0&0&0&0&0&g&{{E_{ex}}}&0&0&{{\Omega _p}}&0\\
0&0&0&0&0&{{\Omega _p}}&0&{{\omega _1} + {\omega _{VUV}}}&0&g&0\\
0&0&0&0&0&0&0&0&{{\omega _2}}&0&0\\
0&0&0&0&0&0&{{\Omega _p}}&g&0&{{\omega _1} + {E_{ex}}}&0\\
0&0&{{\Omega _p}}&0&g&0&0&0&0&0&{{\omega _1} + {\omega _2} + {E_{ex}}}
\end{array}} \right)]
    \end{equation}
\end{widetext}

\section{Derivation and Estimation of the Single-Photon–Nuclear Coupling Strength $g$}

We derive the single-photon–nucleus coupling strength $g$ using cavity QED,  and evaluate its numerical value under experimentally relevant conditions based on the latest parameters of the $^{229}\text{Th}$ nuclear isomer transition.

The isomeric transition of $^{229}\text{Th}$ is of magnetic-dipole (M1) character. Under the rotating-wave approximation (RWA), the single-nucleus-cavity interaction Hamiltonian reads
\begin{equation}
    \hat{H}_{\text{int}} = -\hat{\mu} \cdot \hat{B},
\end{equation}
where $\hat{\mu}$ denotes the nuclear magnetic dipole operator and $\hat{B}$ is the quantized magnetic field operator.

The single-photon coupling strength $g$ is defined as half of the vacuum Rabi frequency and can be written as
\begin{equation}
    g = \frac{|\mu_{ge}| B_{\text{vac}}}{\hbar},
    \label{eq:g_def}
\end{equation}
where $|\mu_\mathrm{ge}|$ is the transition magnetic dipole moment and $B_\mathrm{vac}$ is the vacuum magnetic field amplitude for a cavity mode of volume $V_\mathrm{eff}$. 

The transition matrix element is
\begin{equation}
M_{\mathbf k\lambda}
=
\langle G,1_{\mathbf k\lambda}|\hat H_{\rm int}|E,0\rangle
=
-i
\sqrt{\frac{\mu_0\hbar\omega_k}{2V}}
\,
\boldsymbol{\mu}_{ge}\cdot(\hat{\mathbf k}\times\boldsymbol{\epsilon}_{\mathbf k\lambda}),
\end{equation}
where $\hat{\mathbf k}$ is wavevector, $\boldsymbol{\epsilon}_{\mathbf k\lambda}$ denotes the unit polarization vector of the transverse electromagnetic mode, satisfying $\boldsymbol{\epsilon}_{\mathbf k\lambda}\cdot\mathbf k=0$ and $\boldsymbol{\epsilon}_{\mathbf k\lambda}\cdot \boldsymbol{\epsilon}_{\mathbf k\lambda'}=\delta_{\lambda\lambda'}$. Using Fermi's golden rule, the spontaneous emission rate is
\begin{equation}
\Gamma
=
\frac{2\pi}{\hbar}
\sum_{\mathbf k,\lambda}
|M_{\mathbf k\lambda}|^2
\delta(\hbar\omega_{\rm nuc}-\hbar\omega_k).
\end{equation}
Replacing the summation over modes by an integral
\begin{equation}
\sum_{\mathbf k}
\rightarrow
\frac{V}{(2\pi)^3}\int d^3k ,
\end{equation}
and using $\omega_k=ck$, one obtains
\begin{equation}
\Gamma
=
\frac{\mu_0}{8\pi^2}
\frac{\omega_{\rm nuc}^3}{c^3}
\sum_\lambda
\int d\Omega
\left|
\boldsymbol{\mu}_{ge}\cdot(\hat{\mathbf k}\times\boldsymbol{\epsilon}_{\mathbf k\lambda})
\right|^2 .
\end{equation}
Summing over the two transverse polarizations gives
\begin{equation}
\sum_\lambda
\left|
\boldsymbol{\mu}_{ge}\cdot(\hat{\mathbf k}\times\boldsymbol{\epsilon}_{\mathbf k\lambda})
\right|^2
=
|\boldsymbol{\mu}_{ge}|^2
-
|\boldsymbol{\mu}_{ge}\cdot\hat{\mathbf k}|^2 .
\end{equation}
Performing the angular integration yields
\begin{equation}
\int d\Omega
\left(
|\boldsymbol{\mu}_{ge}|^2
-
|\boldsymbol{\mu}_{ge}\cdot\hat{\mathbf k}|^2
\right)
=
\frac{8\pi}{3}|\boldsymbol{\mu}_{ge}|^2 .
\end{equation}
Finally, the magnetic-dipole spontaneous emission rate becomes
\begin{equation}
\Gamma_{\rm M1}
=
\frac{\mu_0\omega_{\rm nuc}^3}{3\pi\hbar c^3}
|\mu_{ge}|^2 .
\label{eq:mu_solved}
\end{equation}


The vacuum magnetic field $B_{\text{vac}}$ is determined by the quantization condition of the cavity mode. Equating the zero-point magnetic energy to $\frac{1}{4}\hbar\omega_\mathrm{nuc}$ yields 
\begin{equation}
    \int \frac{|B|^2}{2\mu_0} \, dV = \frac{1}{4} \hbar \omega_\text{nuc}
    \;\;\Rightarrow\;\;
    \frac{B_{\text{vac}}^2 V_{\text{eff}}}{2\mu_0} = \frac{1}{4} \hbar \omega_\text{nuc}.
\end{equation}
The vacuum magnetic field amplitude is therefore 
\begin{equation}
    B_{\text{vac}} = \sqrt{\frac{\mu_0 \hbar \omega_\text{nuc}}{2 V_{\text{eff}}}}.
    \label{eq:B_vac}
\end{equation}

Substituting Eqs.~(\ref{eq:mu_solved}) and (\ref{eq:B_vac}) into Eq.~(\ref{eq:g_def}), we obtain an analytical expression for $g$ that depends only on measurable physical quantities $\Gamma_{\text{vac}}$, $\omega_\text{nuc}$, and the mode volume $V_{\text{eff}}$:
\begin{equation}
    g = \frac{1}{\hbar}
    \left( \sqrt{\frac{3\pi \hbar c^3 \Gamma_{\text{vac}}}{\mu_0 \omega_\text{nuc}^3}} \right)
    \left( \sqrt{\frac{\mu_0 \hbar \omega_\text{nuc}}{2 V_{\text{eff}}}} \right)
    = \sqrt{\frac{3\pi c^3 \Gamma_{\text{vac}}}{2 \omega_\text{nuc}^2 V_{\text{eff}}}}.
    \label{eq:g_final}
\end{equation}

We evaluate $g$ using the basic parameters of the nuclear transition: transition wavelength $\lambda = 148.3821~\text{nm}$, transition frequency $\omega_{\text{nuc}} \approx 1.269 \times 10^{16}~\text{rad/s}$, and an intrinsic vacuum lifetime $\tau_{\text{vac}} = 1740(50)~\text{s}$ (corrected for refractive-index effects, yielding an inferred magnetic dipole moment of $|\mu_{\text{ge}}| \approx 0.48 \mu_N$).

To substantially enhance the single-photon-nucleus coupling strength and access the strong-coupling regime, we introduce a high-finesse optical microcavity external to the crystal. This confines the electromagnetic field to a micrometer-scale volume. For the interaction region, we assume a cavity length $L = 5~\mu\text{m}$ and a beam waist radius $w_0 = 8.5~\mu\text{m}$, yielding a drastically compressed effective mode volume of $V_{\text{eff}} \approx \pi w_0^2 L \approx 1.0 \times 10^{-15}~\text{m}^3$.

Compared with current macroscopic experimental setups, this design reduces the effective mode volume by approximately three orders of magnitude. Substituting these parameters into Eq.~(\ref{eq:g_final}), we obtain the microcavity-enhanced single-photon–nucleus coupling strength
\begin{equation}
    g_{\text{micro}} \approx {671.13}\  \rm {rad/s}
    \quad (\text{or } \sim {106.8}\ \rm {Hz}).
\end{equation}

\section{Magnetic-dipole operator and angular-momentum selection rules}
The nuclear magnetic dipole operator originates from the orbital and spin angular momenta of the nucleons, which can be written as
\begin{equation}
\hat{\boldsymbol{\mu}}
=
\mu_N
\sum_i
\left(
g_l^{(i)} \hat{\mathbf l}_i
+
g_s^{(i)} \hat{\mathbf s}_i
\right),
\end{equation}
where
$\mu_N = e\hbar/(2m_p)$ is the nuclear magneton, $i$ labels the nucleons inside the nucleus, $\hat{\mathbf l}_i$ is the orbital angular-momentum operator of nucleon $i$, $\hat{\mathbf s}_i$ is the spin angular-momentum operator of nucleon $i$, $g_l^{(i)}$ and $g_s^{(i)}$ are the orbital and spin $g$-factors.

Because angular momentum operators transform as vectors under rotation operations, the magnetic dipole operator $\hat{\boldsymbol{\mu}}$ forms an
irreducible spherical rank-1 tensor operator. Hence, it is convenient to express the components $(\mu_x,\mu_y,\mu_z)$ in the spherical basis

\begin{equation}
\mu_0 = \mu_z ,
\end{equation}

\begin{equation}
\mu_{\pm1} =
\mp \frac{1}{\sqrt{2}}
\left(
\mu_x \pm i \mu_y
\right).
\end{equation}
These components satisfy the defining commutation relations of a rank-1 spherical tensor with respect to the total angular-momentum operator
$\hat{\mathbf J}$,

\begin{equation}
[\hat J_z , \mu_q] = \hbar q \mu_q ,
\end{equation}

\begin{equation}
[\hat J_{\pm} , \mu_q]
=
\hbar
\sqrt{(1 \mp q)(1 \pm q + 1)}
\,\mu_{q\pm1},
\end{equation}
where $\hat{\mathbf J}$ is the total nuclear angular-momentum operator, $\hat J_z$ is its $z$-component, $\hat J_{\pm} = \hat J_x \pm i\hat J_y$ are the angular-momentum ladder operators, $q = 0,\pm1$ labels the spherical components of the tensor operator.

Because $\hat{\boldsymbol{\mu}}$ is a rank-1 tensor operator, its matrix elements between different nuclear states can be expressed using the Wigner-Eckart theorem,

\begin{equation}
\langle \alpha_f J_f M_f | \mu_q | \alpha_i J_i M_i \rangle
=
\frac{
\langle \alpha_f J_f \Vert \mu \Vert \alpha_i J_i \rangle
}{
\sqrt{2J_f+1}
}
C^{J_f M_f}_{J_i M_i\,1 q},
\end{equation}
where $J_i$ and $J_f$ are the total angular momenta of the initial and final nuclear states, $M_i$ and $M_f$ are the corresponding magnetic quantum numbers, $\alpha_i$ and $\alpha_f$ denote additional nuclear quantum numbers that specify the nuclear configuration, $C^{J_f M_f}_{J_i M_i\,1 q}$ is the Clebsch-Gordan coefficient describing the coupling of angular momenta, $\langle \alpha_f J_f \Vert \mu \Vert \alpha_i J_i \rangle$ is the reduced magnetic dipole matrix element.

The Clebsch-Gordan coefficient is nonzero only if the triangle condition for angular-momentum coupling is satisfied, i.e.,
\begin{equation}
|J_i - 1| \le J_f \le J_i + 1.
\end{equation}
Therefore a magnetic-dipole operator can connect nuclear states with
\begin{equation}
\Delta J = J_f - J_i = 0, \pm1.
\end{equation}
For the $^{229}$Th nuclear isomer transition,
the quantum numbers are
\begin{equation}
J_e^\pi = 3/2^+,
\qquad
J_g^\pi = 5/2^+,
\end{equation}
which gives
\begin{equation}
\Delta J = 1,
\end{equation}
with the parity of the nuclear state remaining the same. Thus the transition satisfies the M1 selection rules and is an allowed magnetic-dipole transition. The light-matter coupling strength depends on the transition magnetic dipole matrix element
\begin{equation}
\mu_{ge}
=
\langle g | \hat{\boldsymbol{\mu}} | e \rangle ,
\end{equation}
which corresponds to the reduced magnetic dipole matrix element between the nuclear ground state $|g\rangle$ and the isomeric state $|e\rangle$.

\section{Derivation of the Maxwell-Bloch Equations}
Based on the Heisenberg equations and the mean-field approximation, we derived the equations of motion for the coupled $^{229}$Th system.

For an open quantum system described by the density operator $\rho$, the time evolution of the expectation value of an arbitrary system operator $\hat{O}$ is given by
\begin{equation}
\frac{d}{dt}\langle \hat{O} \rangle = -i \langle [\hat{O}, \hat{H}] \rangle + \mathrm{Tr}(\hat{O} \mathcal{L}[\rho]),
\label{eq:A1}
\end{equation}
where we set $\hbar = 1$, $\hat{H}$ is the total system Hamiltonian, and $\mathcal{L}[\rho]$ represents the Lindblad dissipator describing interactions with the environment. The Hamiltonian in the rotating frame, $H = H_0 + H_{\rm int} + H_{\rm FWM}$, is defined in the main text. In the following derivation, we assume the system is at resonance ($\Delta_{\rm VUV} \approx 0, \Delta_{\rm nuc} \approx 0$).

\subsection{VUV Field Equation}

For the VUV cavity field amplitude $\alpha(t) = \langle \hat{a}_{\mathrm{VUV}} \rangle$ (hereafter denoted as $\hat{a}$), setting $\hat{O} = \hat{a}$ in Eq.~(\ref{eq:A1}) yields:

The relevant interaction terms are $H_{\rm int} = g(\hat{a}^\dagger \hat{J}_- + \hat{a} \hat{J}_+)$ and $H_{\rm FWM} = U(\hat{a}^\dagger \hat{a}_2^\dagger \hat{a}_1^2 + {\rm h.c.})$. Using bosonic commutation relations $[\hat{a}, \hat{a}^\dagger] = 1$:
\begin{equation}
[\hat{a}, \hat{H}] = [\hat{a}, H_{\rm int}] + [\hat{a}, H_{\rm FWM}].
\end{equation}
The interaction contribution reads
\begin{equation}
[\hat{a}, H_{\rm int}] = g [\hat{a}, \hat{a}^\dagger \hat{J}_-] = g \hat{J}_-.
\end{equation}
The FWM term contributes
\begin{equation}
[\hat{a}, H_{\rm FWM}] = U [\hat{a}, \hat{a}^\dagger \hat{a}_2^\dagger \hat{a}_1^2] = U \hat{a}_2^\dagger \hat{a}_1^2.
\end{equation}
The nonlinear source term $U \hat{a}_2^\dagger \hat{a}_1^2$ converts two pump photons ($\omega_1$) into one seed photon ($\omega_2$) and one VUV photon.

The cavity decay is described by $\mathcal{L}_{\rm cav}\rho = \kappa \left( \hat{a}\rho\hat{a}^\dagger - \frac{1}{2}\{\hat{a}^\dagger \hat{a}, \rho\} \right)$. Its contribution to the expectation value is
\begin{equation}
\mathrm{Tr}(\hat{a} \mathcal{L}_{\rm cav}\rho) = -\frac{\kappa}{2} \langle \hat{a} \rangle.
\end{equation}

Combining the above, we obtain
\begin{equation}
\frac{d}{dt} \langle \hat{a} \rangle = -i g \langle \hat{J}_- \rangle - i U \langle \hat{a}_2^\dagger \hat{a}_1^2 \rangle - \frac{\kappa}{2} \langle \hat{a} \rangle.
\end{equation}

For $N$ identical nuclei, the collective spin operator decomposes as $\langle \hat{J}_- \rangle = \sum_k \langle \hat{\sigma}_-^{(k)} \rangle = N P$, with the single-nucleus polarization $P = \langle \hat{\sigma}_- \rangle$. Treating the strong pump and seed fields classically, we define the FWM source as $\mathcal{E}(t) = -i U \langle \hat{a}_2^\dagger \rangle \langle \hat{a}_1 \rangle^2$. Substituting these recovers Eq. (10) of the main text:
\begin{equation}
\dot{\alpha} = \mathcal{E}(t) - \frac{\kappa}{2} \alpha - i N g P.
\end{equation}

\subsection{Nuclear Polarization Equation}

We derive the equation for the microscopic nuclear polarization $P(t) = \langle \hat{\sigma}_- \rangle$.

The relevant Hamiltonian for the $j$-th nucleus is $H_{\rm int}^{(j)} = g(\hat{a}^\dagger \hat{\sigma}_- + \hat{a} \hat{\sigma}_+)$. Using Pauli commutation relations $[\hat{\sigma}_-, \hat{\sigma}_+] = -\hat{\sigma}_z$:
\begin{equation}
[\hat{\sigma}_-, \hat{H}] = g [\hat{\sigma}_-, \hat{a} \hat{\sigma}_+] = g \hat{a} [\hat{\sigma}_-, \hat{\sigma}_+] = - g \hat{a} \hat{\sigma}_z.
\end{equation}

The nuclear decay is modeled by $\mathcal{L}_{\rm nuc}\rho = \gamma_- \left( \hat{\sigma}_- \rho \hat{\sigma}_+ - \frac{1}{2} \{ \hat{\sigma}_+ \hat{\sigma}_-, \rho \} \right)$. Its effect on the polarization is
\begin{equation}
\mathrm{Tr}(\hat{\sigma}_- \mathcal{L}_{\rm nuc} \rho) = -\frac{\gamma_-}{2} \langle \hat{\sigma}_- \rangle.
\end{equation}

The equation of motion becomes
\begin{equation}
\frac{d}{dt} \langle \hat{\sigma}_- \rangle = - i \langle - g \hat{a} \hat{\sigma}_z \rangle - \frac{\gamma_-}{2} \langle \hat{\sigma}_- \rangle.
\end{equation}
Decoupling quantum correlations $\langle \hat{a} \hat{\sigma}_z \rangle \approx \langle \hat{a} \rangle \langle \hat{\sigma}_z \rangle = \alpha Z$, we obtain Eq. (11) of the main text:
\begin{equation}
\dot{P} = -\frac{\gamma_-}{2} P + i g \alpha Z.
\end{equation}

\subsection{Population Inversion Equation}

We derive the equation for the population inversion $Z(t) = \langle \hat{\sigma}_z \rangle$.

Using $[\hat{\sigma}_z, \hat{\sigma}_-] = -2 \hat{\sigma}_-$ and $[\hat{\sigma}_z, \hat{\sigma}_+] = 2 \hat{\sigma}_+$:
\begin{align}
[\hat{\sigma}_z, \hat{H}] &= g [\hat{\sigma}_z, \hat{a}^\dagger \hat{\sigma}_- + \hat{a} \hat{\sigma}_+] \nonumber \\
&= g \hat{a}^\dagger [\hat{\sigma}_z, \hat{\sigma}_-] + g \hat{a} [\hat{\sigma}_z, \hat{\sigma}_+] \nonumber \\
&= -2 g \hat{a}^\dagger \hat{\sigma}_- + 2 g \hat{a} \hat{\sigma}_+.
\end{align}

The population inversion decay term drives the system toward the ground state:
\begin{equation}
\mathrm{Tr}(\hat{\sigma}_z \mathcal{L}_{\rm nuc} \rho) = -\gamma_- ( \langle \hat{\sigma}_z \rangle + 1 ),
\end{equation}
ensuring $Z \to -1$ in the absence of driving.

The equation of motion reads
\begin{equation}
\frac{d}{dt} \langle \hat{\sigma}_z \rangle = - i \langle -2 g \hat{a}^\dagger \hat{\sigma}_- + 2 g \hat{a} \hat{\sigma}_+ \rangle - \gamma_- (Z + 1),
\end{equation}
which can be rearranged as
\begin{align}
\dot{Z} &= 2 i g [ \langle \hat{a}^\dagger \hat{\sigma}_- \rangle - \langle \hat{a} \hat{\sigma}_+ \rangle ] - \gamma_- (Z + 1).
\end{align}
Using $\langle \hat{a}^\dagger \hat{\sigma}_- \rangle \approx \alpha^* P$ and $\langle \hat{a} \hat{\sigma}_+ \rangle \approx \alpha P^*$, we obtain Eq. (12) of the main text:
\begin{equation}
\dot{Z} = - \gamma_- (Z + 1) + 2 i g (\alpha^* P - \alpha P^*).
\end{equation}
Note that $(\alpha^* P)^* = \alpha P^*$, so the term in parentheses is purely imaginary, ensuring $\dot{Z}$ is real. Explicitly, $2 i g (\alpha^* P - \alpha P^*) = -4 g \, \mathrm{Im}[\alpha^* P]$.

\section{Theoretical Verification: $\sqrt{N}$ Scaling from Linearized Analysis}
\label{app:linear_analysis}

To confirm the collective strong-coupling regime and to explain the $\sqrt{N}$ enhancement observed in numerical simulations, we perform a linear stability analysis of the semiclassical Maxwell-Bloch equations.

We consider the system in the weak-excitation and strong-coupling limits, setting the external driving $\mathcal{E} = 0$ to study the free evolution, and $\kappa = \gamma_- = 0$ to neglect dissipation for extracting the eigenfrequencies. Furthermore, we assume the system remains predominantly in the ground state, i.e., $Z(t) \approx -1$.

Under these assumptions, the polarization equation $\dot{P} = i g \alpha Z$ simplifies to $\dot{P} = - i g \alpha$, and the dynamics are governed by the linear coupled equations:
\begin{equation}
\begin{cases}
\frac{d \alpha}{dt} = - i (N g) P, \\[2mm]
\frac{d P}{dt} = - i g \alpha.
\end{cases}
\label{eq:linear_system}
\end{equation}

Combining the two equations gives
\begin{equation}
\frac{d^2 \alpha}{dt^2} = - i N g \frac{dP}{dt} = - i N g (- i g \alpha).
\end{equation}
which reduces to a standard harmonic oscillator equation: 
\begin{equation}
\frac{d^2 \alpha}{dt^2} = - (N g^2) \alpha.
\label{eq:harmonic_oscillator}
\end{equation}
Comparing with $\ddot{x} + \omega^2 x = 0$, the system eigenfrequency (vacuum Rabi frequency $\omega_{\rm VR}$) is
\begin{equation}
\omega_{\rm VR} = \sqrt{N} g.
\end{equation}

This analytical result demonstrates the $\sqrt{N}$ scaling of the collective coupling. As the number of interacting nuclei $N$ increases, the effective light–matter coupling is significantly enhanced, thereby overcoming the limitation of the weak single-particle coupling $g$ and enabling the realization of a strong-coupling regime.

\section{Derivation of Hopfield Coefficients for Nuclear Polaritons}
\label{app:hopfield}

In this Appendix, we present the explicit derivation of the Hopfield coefficients characterizing the hybridization between the VUV cavity field and the collective nuclear excitation. These coefficients quantify the photonic and nuclear components of the polaritonic eigenstates and provide a rigorous criterion for the formation of nuclear polaritons.

\subsection{Effective Hamiltonian in the Single-Excitation Subspace}

Under the conditions considered in this work, the relevant dynamics of the coupled system occur within the single-excitation subspace spanned by the bare cavity photon state $\ket{G,1_{\mathrm{VUV}}}$ and the collective nuclear excited state $\ket{E,0}$. In the rotating frame with respect to the nuclear transition frequency, the effective Hamiltonian takes the form

\begin{equation}\label{eq:Heff}
H_\text{eff}=   \begin{pmatrix}
\Delta & g \\ g & 0
\end{pmatrix}
\end{equation}
where $\Delta = \omega_{\mathrm{VUV}} - \omega_\mathrm{nuc}$ denotes the cavity-nuclear detuning and $g$ is the collective vacuum Rabi coupling strength.

\subsection{Polariton Eigenstates}

The eigenvalue problem $H_{\mathrm{eff}}\ket{\Psi} = E \ket{\Psi}$ yields two eigenenergies,
\begin{equation}
E_{\pm}
=
\frac{\Delta}{2}
\pm
\frac{1}{2}\sqrt{\Delta^{2} + 4 g^{2}},
\label{eq:polariton_energies}
\end{equation}
corresponding to the upper ($+$) and lower ($-$) polariton branches.

The corresponding normalized eigenstates can be written as
\begin{align}
\ket{\mathrm{LP}}=C_{\mathrm{LP}} \ket{G,1_{\mathrm{VUV}}}-X_{\mathrm{LP}} \ket{E,0},\\
\ket{\mathrm{UP}}=C_{\mathrm{UP}} \ket{G,1_{\mathrm{VUV}}}+X_{\mathrm{UP}} \ket{E,0},
\label{eq:polariton_states}
\end{align}
where $C$ and $X$ denote the photonic and nuclear amplitudes, respectively.

\subsection{Hopfield Coefficients}

Imposing the eigenvalue equations together with the normalization condition
$|C|^{2} + |X|^{2} = 1$, one obtains the Hopfield coefficients in closed analytical form. For the lower polariton branch, they read
\begin{align}
|C_{\mathrm{LP}}|^{2}
&=
\frac{1}{2}
\left(
1 - \frac{\Delta}{\sqrt{\Delta^{2} + 4 g^{2}}}
\right),
\\
|X_{\mathrm{LP}}|^{2}
&=
\frac{1}{2}
\left(
1 + \frac{\Delta}{\sqrt{\Delta^{2} + 4 g^{2}}}
\right).
\label{eq:hopfield_lp}
\end{align}
The Hopfield coefficients for the upper polariton are obtained by exchanging the signs in Eq.\eqref{eq:hopfield_lp}.

At zero detuning ($\Delta = 0$), the coefficients satisfy $|C|^{2} = |X|^{2} = 1/2$, indicating maximal light-matter hybridization. In the limits $|\Delta| \gg g$, the eigenstates recover their bare cavity photon or nuclear excitation character, respectively. These features provide a clear and quantitative signature of nuclear polariton formation.

\section{Theory and Simulation of Collective Quantum Effects Driven by Four-Wave Mixing}

\subsection{Rotating-Wave Approximation}

We define the unitary transformation:
\begin{multline}
W(t) = \exp \Bigl[ -i \bigl( \omega_1 a_1^\dagger a_1 + \omega_2 a_2^\dagger a_2 \\
+ \omega_{\mathrm{VUV}} a_{\mathrm{VUV}}^\dagger a_{\mathrm{VUV}} + \omega_{\mathrm{ex}} J_z \bigr) t \Bigr],
\end{multline}
where $\omega_i$ are the central frequencies of the optical modes and $\omega_{\rm ex}$ is the atomic transition frequency, assumed near-resonant with the VUV mode, i.e., $\omega_{\rm ex} \approx \omega_{\rm VUV}$.

The transformed Hamiltonian is
\begin{equation}
\tilde{H} = W^\dagger(t) H W(t) - i W^\dagger(t) \dot{W}(t).
\end{equation}

The time derivative term $- i W^\dagger(t) \dot{W}(t)$ generates the detunings $\Delta_k = \omega^{(j)}_k - \omega_k$ and $\Delta_{\rm ex} = E_{\rm ex} - \omega_{\rm ex}$, yielding the free Hamiltonian in the rotating frame:
\begin{equation}
\tilde{H}_0 = \Delta_1 a_1^\dagger a_1 + \Delta_2 a_2^\dagger a_2 + \Delta_{\rm VUV} a_{\rm VUV}^\dagger a_{\rm VUV} + \Delta_{\rm ex} J_z.
\end{equation}

Transforming the Dicke interaction into the rotating frame introduces time-dependent phases. Near resonance ($\omega_{\rm VUV} \approx \omega_{\rm ex}$), we apply the RWA to neglect the rapidly oscillating counter-rotating terms ($a_{\rm VUV}^\dagger J_+$ and $a_{\rm VUV} J_-$). The interaction thus simplifies to the standard Tavis-Cummings form:
\begin{equation}
\tilde{H}_{\rm int} = g \left( a_{\rm VUV}^\dagger J_- + a_{\rm VUV} J_+ \right).
\end{equation}

Consider $a_{\rm VUV}^\dagger a_2^\dagger a_1 a_1$:
\begin{equation}
\tilde{H}_{\rm FWM} \propto a_{\rm VUV}^\dagger a_2^\dagger a_1 a_1 e^{i (\omega_{\rm VUV} + \omega_2 - 2 \omega_1) t}.
\end{equation}
Energy conservation requires
\begin{equation}
2 \omega_1 = \omega_{\rm VUV} + \omega_2,
\end{equation}
so the term remains in the RWA while non-matching terms are discarded.

The resulting Hamiltonian in the rotating frame is:
\begin{equation}
\begin{aligned}
H_{\rm RWA} &= \Delta_1 a_1^\dagger a_1 + \Delta_2 a_2^\dagger a_2 + \Delta_{\rm VUV} a_{\rm VUV}^\dagger a_{\rm VUV} + \Delta_{\rm ex} J_z \\
&\quad + g ( a_{\rm VUV}^\dagger J_- + a_{\rm VUV} J_+ ) \\
&\quad + U ( a_{\rm VUV}^\dagger a_2^\dagger a_1 a_1 + \text{h.c.} ) \\
&\quad + \Omega_p e^{-(t-t_0)^2 / 2\sigma^2} (a_1 + a_1^\dagger).
\end{aligned}
\end{equation}

\subsection{Master Equation with Dissipation}

The model incorporates the relevant dissipation channels, characterized by the optical cavity decay rates ($\kappa_1$, $\kappa_2$, and $\kappa_\mathrm{VUV}$) and the collective atomic decay rate ($\gamma_-$). 

The Lindblad master equation is
\begin{equation}
\begin{split}
\dot{\rho} =& -i[H_{\rm RWA}, \rho] + \sum_{j=1,2} \kappa_j \mathcal{D}[a_j]\rho \\&+ \kappa_{\rm VUV} \mathcal{D}[a_{\rm VUV}]\rho \\&+ \gamma_- \mathcal{D}[J_-]\rho,
\end{split}
\end{equation}
with
\begin{equation}
\mathcal{D}[O]\rho = O \rho O^\dagger - \frac{1}{2} \{O^\dagger O, \rho\}.
\end{equation}

In the bad-cavity limit ($\kappa_{\rm VUV} \gg g, U$), the Heisenberg equation for $a_{\rm VUV}$ reads:
\begin{equation}
\dot{a}_{\rm VUV} = - \left( \frac{\kappa_{\rm VUV}}{2} + i \Delta_{\rm VUV} \right) a_{\rm VUV} - i \hat{B} + \xi(t),
\end{equation}
where $\hat{B} = g J_- + U a_2^\dagger a_1^2$ is the collective source operator. Setting $\dot{a}_{\rm VUV} \approx 0$ and neglecting noise, we obtain
\begin{equation}
a_{\rm VUV} \approx \frac{-i}{\mathcal{L}} \hat{B}, \quad \mathcal{L} = \frac{\kappa_{\rm VUV}}{2} + i \Delta_{\rm VUV}.
\end{equation}

Substituting into $H_{\rm int} = a_{\rm VUV}^\dagger \hat{B} + \text{h.c.}$ yields the Hamiltonian correction:
\begin{equation}
H_{\rm add} = - \frac{2 \Delta_{\rm VUV}}{ (\kappa_{\rm VUV}/2)^2 + \Delta_{\rm VUV}^2 } \hat{B}^\dagger \hat{B}.
\end{equation}
On resonance ($\Delta_{\rm VUV}=0$), $H_{\rm add}=0$.
Similarly, the effective dissipation is
\begin{equation}
\mathcal{L}_{\rm eff}\rho = \frac{\kappa_{\rm VUV}}{ (\kappa_{\rm VUV}/2)^2 + \Delta_{\rm VUV}^2 } \mathcal{D}[\hat{B}] \rho,
\end{equation}
reducing to $4/\kappa_{\rm VUV}$ on resonance. The reduced system evolves according to
\begin{equation}
\begin{split}
\dot{\rho} = &-i [H_{\rm sys}, \rho] + \sum_{j=1,2} \kappa_j \mathcal{D}[a_j]\rho \\
&+ \gamma_- \mathcal{D}[J_-]\rho \\
&+ \frac{4}{\kappa_{\rm VUV}} \mathcal{D}[g J_- + U a_2^\dagger a_1^2]\rho.
\end{split}
\end{equation}

For a Lindblad operator $\mathcal{D}[\hat{A}+\hat{B}]$ with $\hat{B}$ a strong coherent field (c-number), the cross terms can be exactly rewritten as a unitary contribution:
\begin{equation}
\mathcal{L}_{\rm cross} \rho = -i [H_{\rm eff}, \rho], \quad
H_{\rm eff} = i \frac{\gamma_-}{2} (\hat{B}^\dagger \hat{A} - \hat{B} \hat{A}^\dagger ).
\end{equation}
Applying this framework to the FWM system, we identify the effective decay rate  $\gamma_- = 4/\kappa_{\rm VUV}$, and define the operators $\hat{A} = g J_-$and $\hat{B} = U a_2^\dagger a_1^2$.
\begin{equation}
H_{\rm eff}^{\rm drive} = \frac{2 g U}{\kappa_{\rm VUV}} \left( -i J_+ a_2^\dagger a_1^2 + \text{h.c.} \right),
\end{equation}
describing a coherent drive of the atoms mediated by virtual VUV photons.

The remaining diagonal terms yield the Purcell-enhanced collective decay $\frac{4 g^2}{\kappa_{\rm VUV}} \mathcal{D}[J_-]$, which is retained in the effective dynamics, while the higher-order photonic loss $\frac{4 U^2}{\kappa_{\rm VUV}} \mathcal{D}[a_2^\dagger a_1^2]$ is neglected under the undepleted pump approximation.

\subsection{Physical Interpretation}

In the bad-cavity limit, the VUV mode acts as a virtual level that mediates interactions without storing energy. This unifies the system's dynamics: the coherent FWM input serves as an effective Hamiltonian drive, whereas the collective nuclear emission appears as irreversible dissipation within the same collective channel.

The effective Hamiltonian after eliminating $a_{\rm VUV}$ reads:
\begin{equation}
\begin{aligned}
H_{\rm eff} &= \Delta_1 a_1^\dagger a_1 + \Delta_2 a_2^\dagger a_2 + \Delta_{\rm ex} J_z \\
&\quad + \frac{2 g U}{\kappa_{\rm VUV}} ( J_+ a_2^\dagger a_1^2 + \text{h.c.} ) \\
&\quad + \Omega_0 e^{-(t-t_0)^2/2\sigma^2} (a_1 + a_1^\dagger).
\end{aligned}
\end{equation}
The final effective master equation is expressed as
\begin{equation}
\dot{\rho} = -i [H_{\rm eff}, \rho] + \kappa_1 \mathcal{D}[a_1]\rho + \kappa_2 \mathcal{D}[a_2]\rho + \gamma_{\rm eff} \mathcal{D}[J_-]\rho,
\end{equation}
with $\gamma_{\rm eff} = \gamma_- + 4 g^2 / \kappa_{\rm VUV}$.

To analyze the interplay between coherent excitation and cooperative decay, we apply adiabatic elimination in the bad-cavity limit ($\kappa_{\rm VUV} \gg g\sqrt{N}$). As derived above, the effective nuclear dynamics are governed by two distinct mechanisms: a Purcell-enhanced collective dissipation channel and a phase-sensitive FWM drive mediated by virtual VUV photons. This effective model allows us to manipulate the nuclear lifetime through controlled superradiance.

\begin{figure*}
    \centering
    \includegraphics[width=1\linewidth]{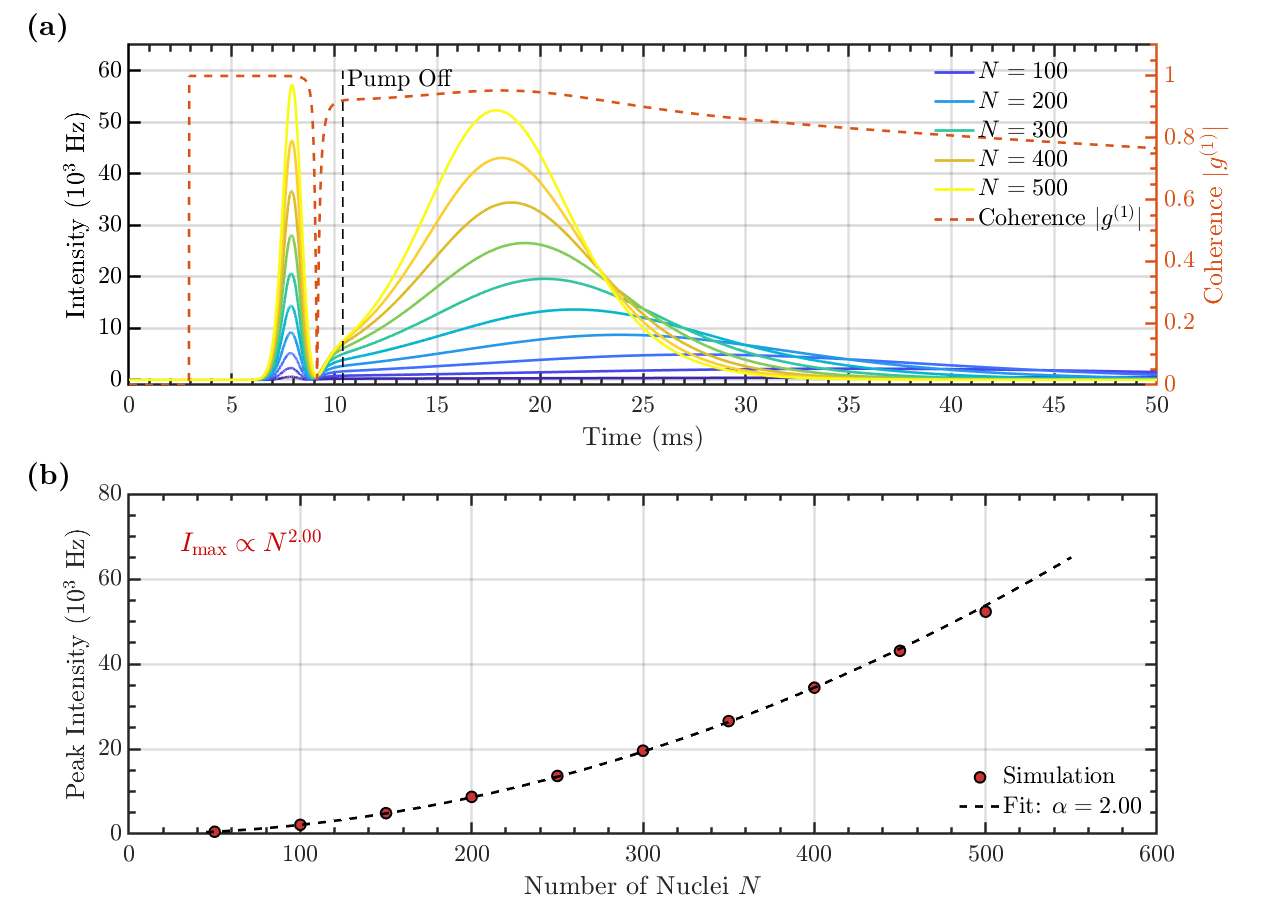}
    \caption{Collective dynamics and scaling of $^{229}\mathrm{Th}$ nuclear superradiance. (a) Time evolution of the radiated intensity $I(t)$ for ensemble sizes $N=100$, $200$, $300$, $400$ and $500$ (simulations were performed for $N=50$-$500$ in steps of $50$; for clarity, only round-hundred values are shown). The vertical dashed line marks the switch-off of the pump field. The subsequent emission peaks correspond to delayed superradiant bursts, whose delay time decreases with increasing $N$. The orange dashed curve (right axis) shows the magnitude of the first-order coherence $g^{(1)}(t)$ for $N=500$. (b) Peak intensity $I_{\max}$ as a function of $N$. The data follow a power-law scaling $I_{\max}\propto N^{2}$ , confirming Dicke-type superradiant emission.}
    \label{figS1}
\end{figure*}

Fig.~\ref{figS1} shows the temporal evolution of the radiated intensity $I(t) = \Gamma_{\mathrm{eff}} \langle J_+ J_- \rangle$ for different ensemble sizes $N$, illustrating collective enhancement and the $N$-dependent dynamics of the nuclear ensemble. The temporal evolution reveals two emission peaks with distinct physical origins. The first, narrow peak occurs during the pump pulse and arises from the effective coherent drive induced by the FWM process. After the pump is switched off, the system evolves freely and releases the stored excitation in a second, broader emission peak. This peak is strongly enhanced with increasing $N$ and represents the collective superradiant burst. Its position shifts systematically to earlier times as $N$ increases, indicating an accelerated buildup of macroscopic coherence, characteristic of Dicke-type collective emission.

Fig.~\ref{figS1} also displays the first-order coherence function $g^{(1)}(t)
=\frac{|\langle J_{-}(t)\rangle|}{\sqrt{\langle J_{+}(t) J_{-}(t)\rangle}} $ for the largest ensemble size $N=500$. Near the superradiant maximum, $g^{(1)}(t)$ approaches unity, demonstrating macroscopic phase coherence across the ensemble. Fig.~\ref{figS1}(b) shows the scaling of the peak emission intensity, $I_{\max}\propto N^{2}$, consistent with the quadratic scaling expected for ideal Dicke superradiance.

\begin{figure*}
    \centering
    \includegraphics[width=1.0\linewidth]{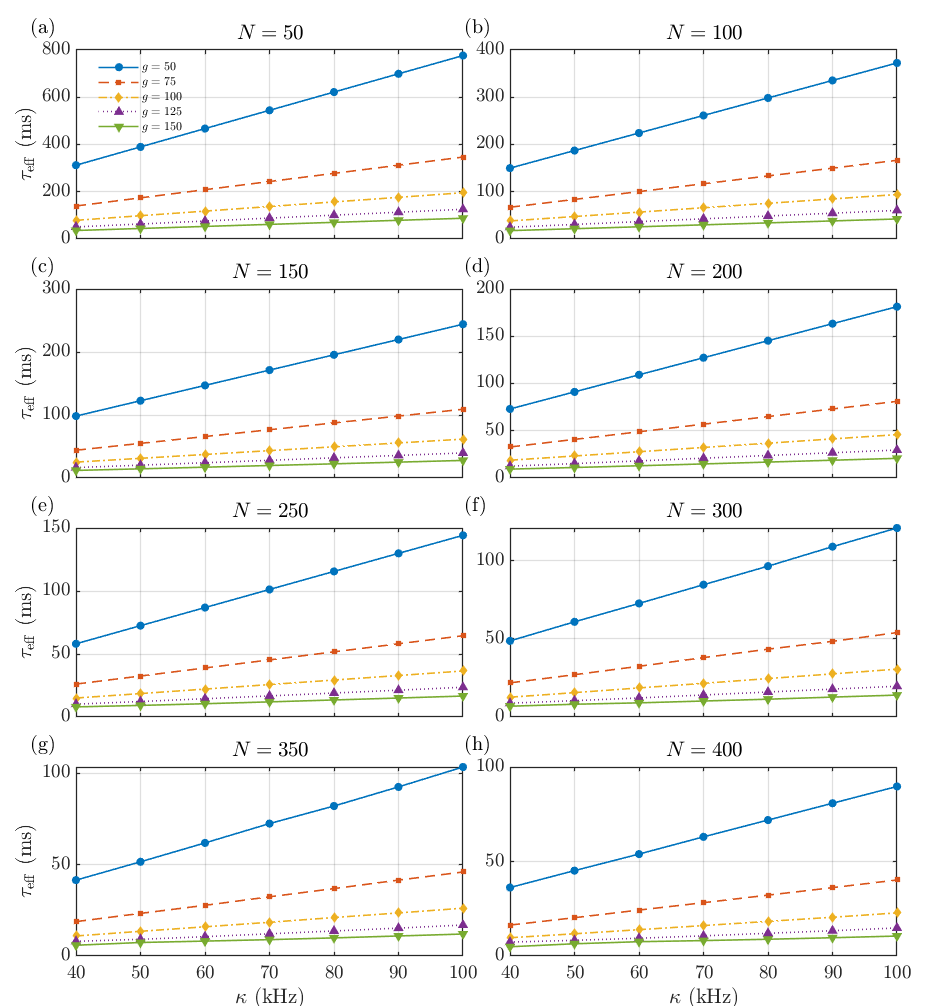}
    \caption{Effective superradiant lifetime $\tau_{\mathrm{eff}}$ versus cavity decay rate $\kappa_{\mathrm{VUV}}$ for ensemble sizes $N=50$-$400$ [(a)-(h)]. Curves correspond to coupling strengths $g\in[50,150]~\mathrm{Hz}$. $\tau_{\mathrm{eff}}$ is extracted from the FWHM of the post-pump emission pulse.}  \label{figS2}
\end{figure*}

Next, we examine  whether the radiative lifetime can be continuously tuned. We characterize it by the temporal width of the superradiant pulse and examine its dependence on cavity dissipation and collective coupling. We quantify this control by analyzing the dependence of the pulse duration $\tau_{\rm eff}$ on the cavity decay rate $\kappa_{\mathrm{VUV}}$. Here, $\tau_\text{eff}$ is defined as the full width at half maximum of the superradiant pulse. As shown in Fig.~\ref{figS2}, $\tau_\text{eff}$ scales linearly with $\kappa_{\mathrm{VUV}}$, confirming the applicability of adiabatic elimination in the bad-cavity regime ($\kappa_{\mathrm{VUV}} \gg g\sqrt{N}  $). This scaling is consistent with the effective decay rate $\Gamma_{\rm eff} \approx 4 N g^2 / \kappa_{\mathrm{VUV}}$, implying $\tau_\mathrm{eff} \sim \kappa_{\mathrm{VUV}} / (N g^2)$. The slope decreases with increasing ensemble size $N$ and coupling strength $g$, directly reflecting cooperative enhancement: the collective nuclear coherence accelerates the radiative decay. The VUV pulse duration can therefore be controlled via cavity parameters, providing a tunable interface between the long-lived nuclear isomer and fast optical modes.

\section{Time-Dependent Hamiltonian}

In the single-excitation subspace, the system is described by the bare-state basis $\{\ket{C}, \ket{N}\}$, representing the VUV cavity photonic state and the collective nuclear excited state, respectively.

Under the strong coupling condition, the time-dependent Hamiltonian of the system can be expressed as:
\begin{equation}
H(t)=
\begin{pmatrix}
\Delta(t) & \Omega \\
\Omega & 0
\end{pmatrix},
\label{eq:H_t}
\end{equation}
where $\Omega = g\sqrt{N}$ is the collective coupling strength. The time-dependent detuning adopts the following sweep protocol:
\begin{equation}
\Delta(t)=\Delta_0 \tanh(kt).
\end{equation}
To ensure asymptotic decoupling of the initial state, the sweep amplitude must satisfy the far-detuned limit $\Delta_0 \gg \Omega$. The parameter $k$ controls the sweep rate, and the sweep velocity at resonance ($t=0$) is $v = \left.d\Delta/dt\right|_{t=0} = \Delta_0 k$.

Diagonalizing $H(t)$ yields the instantaneous eigenenergies of the upper and lower polaritons (UP/LP): 
\begin{equation}
\begin{aligned}
E_{\mathrm{UP}}(t) &= \frac{1}{2}\Big(\Delta(t)+\sqrt{\Delta(t)^2+4\Omega^2}\Big), \\
E_{\mathrm{LP}}(t) &= \frac{1}{2}\Big(\Delta(t)-\sqrt{\Delta(t)^2+4\Omega^2}\Big).
\end{aligned}
\end{equation}

Defining the time-dependent normalization factors:
\begin{equation}
N_{\mathrm{UP/LP}}(t) = \sqrt{\Omega^2+ \big(E_{\mathrm{UP/LP}}(t)-\Delta(t)\big)^2},
\end{equation}
the instantaneous polariton eigenstate column vectors, composed of Hopfield coefficients, are given by:
\begin{equation}
\ket{\mathrm{UP}(t)} = \frac{1}{N_{\mathrm{UP}}(t)} \begin{pmatrix} \Omega \\ E_{\mathrm{UP}}(t)-\Delta(t) \end{pmatrix},
\end{equation}
\begin{equation}
\ket{\mathrm{LP}(t)} = \frac{1}{N_{\mathrm{LP}}(t)} \begin{pmatrix} \Omega \\ E_{\mathrm{LP}}(t)-\Delta(t) \end{pmatrix}.
\end{equation}

In the far-detuned limit ($\Delta_0 \gg \Omega$), the initial polariton states asymptotically decouple into the pure bare states, i.e., $\ket{\mathrm{LP}(-\infty)} \to \ket{C}$ and $\ket{\mathrm{UP}(-\infty)} \to \ket{N}$.

From the analytical expressions of the instantaneous eigenenergies, the UP and LP branches exhibit an avoided crossing at resonance ($t=0$), with a minimum energy gap $\Delta E_{\mathrm{min}} = E_{\mathrm{UP}}(0) - E_{\mathrm{LP}}(0) = 2\Omega$. The dynamical adiabaticity during the sweep is governed by the Landau-Zener parameter
\begin{equation}
\Gamma = \frac{\pi \Omega^2}{k \Delta_0}.
\end{equation}
For $\Gamma \gg 1$, the system adiabatically follows the instantaneous eigenstate, ensuring high-fidelity quantum storage. Conversely, for $\Gamma \lesssim 1$, nonadiabatic Landau-Zener transitions occur between the UP and LP branches near resonance.

\subsection{Dynamical Evolution and Quantum State Projection}

Assuming the state vector of the system at an arbitrary time $t$ is $\ket{\psi(t)} = c_C(t)\ket{C} + c_N(t)\ket{N} = (c_C(t), c_N(t))^\mathrm{T}$, its time evolution is governed by the Schrödinger equation:
\begin{equation}
i\frac{d}{dt} \begin{pmatrix} c_C(t)\\ c_N(t) \end{pmatrix}
= \begin{pmatrix} \Delta_0\tanh(kt) & \Omega\\ \Omega & 0 \end{pmatrix}
\begin{pmatrix} c_C(t)\\ c_nN(t) \end{pmatrix}.
\end{equation}
Choosing the initial time $t = -T_0$ (setting $T_0=5/k$), the system is prepared in the pure photonic initial state: $c_C(-T_0)=1, c_N(-T_0)=0$.

\subsection{Bare-State Population}

To evaluate the storage efficiency, we project the state vector onto the instantaneous bare-state basis. The populations of the upper and lower polariton branches are given by:
\begin{equation}
\begin{aligned}
P_{\mathrm{photon}}(t) &= \left| \braket{C|\psi(t)} \right|^2 = |c_C(t)|^2, \\
P_{\mathrm{nuclear}}(t) &= \left| \braket{N|\psi(t)} \right|^2 = |c_N(t)|^2.
\end{aligned}
\end{equation}

\subsection{Instantaneous Polariton-State Projection}

The polariton populations correspond to the projection of the state onto the instantaneous dressed-state basis. Since the eigenvectors are real-valued, the bra vector is obtained by transposing the column vector.
\begin{equation}
\bra{\mathrm{UP}(t)} = \big( \ket{\mathrm{UP}(t)} \big)^\dagger = \frac{1}{N_{\mathrm{UP}}(t)} \begin{pmatrix} \Omega , E_{\mathrm{UP}}(t)-\Delta(t) \end{pmatrix}.
\end{equation}
Projecting the time-dependent state vector onto this branch yields the rigorous UP population formula:
\begin{equation}
P_{\mathrm{UP}}(t) = \left| \braket{\mathrm{UP}(t)|\psi(t)} \right|^2 
= \left| \frac{\Omega c_c(t) + \big(E_{\mathrm{UP}}(t)-\Delta(t)\big)c_n(t)}{N_{\mathrm{UP}}(t)} \right|^2.
\end{equation}
Similarly, projecting onto the LP branch yields:
\begin{equation}
P_{\mathrm{LP}}(t) = \left| \braket{\mathrm{LP}(t)|\psi(t)} \right|^2 
= \left| \frac{\Omega c_c(t) + \big(E_{\mathrm{LP}}(t)-\Delta(t)\big)c_n(t)}{N_{\mathrm{LP}}(t)} \right|^2.
\end{equation}
Due to the completeness of the eigenbasis, the conservation law $P_{\mathrm{UP}}(t) + P_{\mathrm{LP}}(t) = 1$ is strictly satisfied in numerical calculations by normalization.

\subsection{Scaling Relation of the Transition Time}

To quantitatively characterize the temporal features of the nonadiabatic penetration dynamics, we extract the jump time from the UP population curve:
\begin{equation}
\tau_{\mathrm{jump}} = t_{90} - t_{10},
\end{equation}
where $t_{10}$ and $t_{90}$ mark the times at which $P_{\mathrm{UP}}(t)$ reaches 10\% and 90\% of its asymptotic jump amplitude, respectively. As analyzed in the main text, this transition time window exhibits a strict inverse scaling relation with the sweep rate, namely $\tau_{\mathrm{jump}} \propto k^{-1}$.

\section{SUPPLEMENTARY FIGURES}

\bibliography{apssamp}